# Super-Resolution Simulation for Real-Time Prediction of Urban Micrometeorology

Ryo Onishi[1], Daisuke Sugiyama[1] and Keigo Matsuda[1]

[1]*Center for Earth Information Science and Technology, Japan Agency for Marine-Earth Science and Technology, Yokohama, Japan*

Corresponding author: Ryo Onishi, 3173-25 Showa-machi Kanazawa-ku, Yokohama Kanagawa 236-0001, Japan. E-mail: onishi.ryo@jamstec.go.jp.

**Abstract**

We propose a super-resolution (SR) simulation system that consists of a physics-based meteorological simulation and an SR method based on a deep convolutional neural network (CNN). The CNN is trained using pairs of high-resolution (HR) and low-resolution (LR) images created from meteorological simulation results for different resolutions so that it can map LR simulation images to HR ones. The proposed SR simulation system, which performs LR simulations, can provide HR prediction results in much shorter operating cycles than those required for corresponding HR simulation prediction system. We apply the SR simulation system to urban micrometeorology, which is strongly affected by buildings and human activity. Urban micrometeorology simulations that need to resolve urban buildings are computationally costly and thus cannot be used for operational real-time predictions even when run on supercomputers. We performed HR micrometeorology simulations on a supercomputer to obtain datasets for training the CNN in the SR method. It is shown that the proposed SR method can be used with a spatial scaling factor of 4 and that it outperforms conventional interpolation methods by a large margin. It is also shown that the proposed SR simulation system has the potential to be used for operational urban micrometeorology predictions.

## 1. Introduction

Heat stress is a serious social problem that will become increasingly serious due to the combination of the heat island effect and global warming. According to World Population Prospects 2017 by the United Nations, two-thirds of the population will live in urban cities (urban population: 6.7 billion; total population: 9.8 billion) in 2050. Heat mitigation in urban cities is thus urgently required. Building-resolving heat environment simulations, such as that conducted by Matsuda et al. (2018), are promising for this task. Such simulations have started to be applied for the assessment of real urban areas. They are, however, very computationally expensive and thus cannot be applied for real-time operational predictions even when run on supercomputers.

Real-time operational urban micrometeorology simulations with meter-level spatial resolution could utilize IoT sensor data, which are strongly influenced by buildings and human activity and thus not utilized for conventional weather predictions. Such simulations would facilitate real-time heat mitigation for individuals and urban drone logistics, but their large computational cost must first be reduced.

In this study, we propose a method for speeding up the creation of high-resolution (HR) prediction maps. Super-resolution (SR) downscaling, which maps low-resolution (LR) images to HR ones, with deep learning based on a convolutional neural network (CNN) (LeCun et al., 1989; Vaillant et al., 1994) is utilized. CNNs, which are biologically inspired deep learning models, have attracted attention in computer vision. SR downscaling with a CNN (SRCNN; Dong et al., 2014) outperforms conventional mapping methods such as interpolation. The integration of SR downscaling and LR prediction simulations, referred to as the SR simulation system here, has the potential to produce HR prediction results with a small computational cost. This study demonstrates the robustness of the proposed SR simulation system and discusses the feasibility of applying the system for real-time micrometeorological prediction in urban built-up areas.

## 2. Super-Resolution Simulation System

Figure 1 shows the proposed SR simulation system. HR numerical simulations provide better predictions than those obtained using LR simulations but are more computationally expensive. The SR simulation system consists of an LR simulation and an SR method that maps the resultant LR prediction images to HR ones. This combination provides predictions that are as good as those obtained using the corresponding HR simulation with a much lower computational cost. It should be noted that this study, as a first step, adopts spatial averaging to obtain LR images from HR ones. For more realistic conditions, LR simulations should be performed to obtain LR images and to discuss real applications. This will be done as a next step.

This study focuses on building-resolving urban micro-meteorological simulations that are computationally very expensive and thus cannot be used for operational purposes. The SR simulation system could facilitate operational prediction and contribute to the weather information infrastructure for future smart cities.

**Urban Micrometeorological Simulation**

We use a multiscale atmosphere-ocean coupled model named the Multi-Scale Simulator for the Geoenvironment (MSSG) (Takahashi et al. 2013; Onishi and Takahashi 2012; Sasaki et al. 2016; Matsuda et al. 2018), developed at the Japan Agency for Marine-Earth Science and Technology (JAMSTEC). MSSG covers global, meso-, and urban scales. For urban scales, the atmospheric component of MSSG (i.e., MSSG-A) can be used as a building-resolving large-eddy simulation (LES) model coupled with a three-dimensional radiative transfer model (Matsuda et al. 2018). The governing equations for MSSG-A are the transport equations for compressible flow, which consist of the conservation equations for mass, momentum, and energy, and mixing ratios of water substances including water vapor, liquid and ice cloud particles. HR LESs have been used for heat mitigation at the pedestrian level in urban areas. Example simulation movies are available on National Museum of Engineering Science and Innovation (Miraikan) (the movies are also available at Miraikan's YouTube channel, "MiraikanChannel"). This study performed a huge number of HR LESs with a 5-m resolution to obtain ground truth simulation results. LR results were created by averaging the HR results. Pairs of HR and LR results were used to train the SRCNN.

**Training Data Preparation**

The SRCNN requires HR and LR image pairs for training. We obtained HR results via offline coupling of MSSG-A meteorological simulations (i.e., mesoscale weather simulations) and MSSG-A building-resolving urban micro-meteorological simulations (LESs). The boundary and initial conditions for the LESs are taken from the meteorological simulations covering the urban simulation domains.

Figure 2 shows the computational domains of the present set of simulations. The mesoscale simulations adopted three two-way-coupled nested systems (Fig. 2(a)), with horizontal resolutions of 1 km (domain 1), 300 m (domain 2) and 100 m (domain 3) centered at the urban simulation domain (Fig. 2(b)). For all three nest systems, 55 vertical layers (lowest-layer height: 75 m were used for the 40-km height domain. The boundary and initial conditions for the mesoscale simulations were taken from the Japan Meteorological Agency (JMA) mesoscale analysis data (MANAL).

We performed LESs for a Tokyo metropolitan built-up area shown in Fig. 2(b). The Tokyo domain was centered at 35.680882° N and 139.767019° E and covered a 2 km × 2 km horizontal area with a 5-m horizontal resolution. The domain height was set to 1,500 m and 151 vertical grid points was used. The vertical grid spacing below the height of 500 m was set to 5 m uniformly, while the spacing above was extended continuously. Figure 3 shows the computational domain, which focuses on Tokyo Station. We adopted the same physical schemes for the micrometeorological simulation as described in detail in Matsuda et al. (2018).

To focus on heat mitigation, the LESs were performed for hot summer hours in which the maximum hourly temperature exceeds 35 °C in the years 2013-2017. Each LES was run for each targeted hour. The results from first 10min time integrations were discarded and the rest 50min results were analyzed and used to obtain 1-min-average values. That is, each LES produces 50 sets of 1-min-average data. The 2-m height atmospheric temperatures were visualized and converted into the ground-truth HR image files. The LR images were obtained by spatially-averaging the HR images.

Table 1 shows the image datasets used for training and evaluating the SRCNN. In this study, we performed totally 86 LESs and obtained 4,300 pairs of training images. Four-fifths of the image pairs were used for training, and the rest were used for evaluation. We tested spatial scaling factors of 2 and 4 for SR mapping.

Figure 4 shows an example of a three-dimensional distribution of instantaneous air temperature from the HR simulation. The Volume Data Visualizer for Google Earth (VDVGE; Kawahara, 2012) was utilized for the visualization. The air was warmed up by the land surface, which was heated by solar radiation. The buoyant motion and the turbulent transportation of the warmed air formed the puffy structure of the temperature distribution.

**Convolutional Neural Network (CNN) for Super-Resolution (SR) Downscaling**

SR methods aim to map between LR and HR images (Irani and Peleg, 1991). The mapping in this study is conducted using a deep CNN (LeCun et al., 1989; Vaillant et al., 1994) that outputs an HR image from an LR one. The CNN architecture learns the functional mapping between LR and HR images using three operations, namely patch extraction, non-linear mapping, and reconstruction (Dong et al., 2014).

Figure 5 summarizes the CNN architecture for SR downscaling. Layer 1 consists of 64 filters of 9×9 kernels for patch extraction, Layer 2 consists of 32 filters of 1×1 kernels for non-linear mappings, and the output layer uses 5×5 kernels for reconstruction. The mean square error (MSE) between the original HR image (i.e., the ground-truth) and the HR image reconstructed from the LR one is used to tune the parameters (weights and biases) of the three-layer CNN with rectified linear unit (ReLU) activation for the non-linear operation. Each network was trained using Adam optimization (Kingma and Ba, 2014). Each model was trained using $10^7$ iterations with a batch size of 200. TensorFlow (Abadi et al., 2016), an open-source library for machine learning, was utilized.

## 3.  Results and Discussion
**Training Results**

Figure 6 shows the learning curve of the present CNN for the SR downscaling of a 2-m-height temperature distribution with a spatial scaling factor of 2. The dataset was TKY.x2. The root-mean-square-error (RMSE) decreases and saturates at about 0.16 K after 50 epochs. This means that the training had saturated (i.e., no more improvement could be expected). The peak signal-to-noise ratio (PSNR) is commonly used as a measure of SR downscaling performance: A larger PSNR, corresponding to smaller RMSE, indicates better performance. Figure 6 (b) shows that the PSNR saturated at 41 dB in this training. It took 1.1 hours to complete the learning process shown in Fig. 6 on a single NVIDIA Tesla P100 GPU board (peak performance: 5.3TFLOPS).

Figure 7 shows an example of SR mapping. The HR (5-m-resolution) images are recovered from the LR (20-m-resolution) image. The SRCNN-derived HR image successfully shows small structures (see circled areas in the figure) that are blurred in the LR image. It is confirmed that the SRCNN recovers contrasts in those small structure clearer than the bicubic interpolation, leading to better agreement with the ground-truth.

Table 2 shows the PSNR and RMSE values for the mapping between LR and HR images of the horizontal distribution of 2-m-height temperature. In the table, SRCNN (ImageNet) was trained by using the ImageNet dataset (Krizhevsky et al., 2012) and SRCNN(MSSG) was trained using the present simulation results. It is clearly shown that SRCNN(MSSG) outperforms conventional interpolation methods; including nearest neighbor (NEARNEST), bicubic (BICUBIC) and Lanczos (LANCZOS) (Duchon, 1979). SRCNN(MSSG) slightly outperforms SRCNN(ImageNet), indicating that it was specialized to the building-resolving simulation results.

Even a cutting-edge building-resolving LES such as of Matsuda et al. (2018) has an inherent error of 0.23K in a heat index when a 5-m spatial resolution is used. Liu et al. (2012) compared air temperatures simulated by LESs with high and low spatial resolutions with observed ones, and reported better agreements in RMSE by about 0.2 K when a higher resolution was used. These indicate that a required precision in the SR downscaling would be around 0.2 K. The RMSE values for SRCNNs with spatial scaling factors of 2 and 4 are greatly below and close to this criterion, respectively. This indicates that a spatial scaling factor of 4 is acceptable in terms of mapping accuracy. The computational cost for a simulation with a 4 times coarser spatial resolution is 256 ($=4^4$) times smaller than that for the reference resolution simulation. That is, using SR downscaling with a spatial scaling factor of 4, the HR result can be obtained 256 times faster.

**Computation Time**

 It took about 11 hours on 100 nodes of the Earth Simulator system (each node has a peak performance of 256 GFLOPS), which consists of 5,120 nodes, at JAMSTEC for a 30-min time integration for the HR simulation (i.e., 5-m spatial resolution for a 2 km × 2 km horizontal domain with 151 vertical layers). An estimation from a 10-second time integration computation shows that it would take 980 hours on a modern desktop workstation with an Intel Xeon Silver 4108 1.8-GHz CPU (peak performance: 230 GFLOPS) for a 30-min time integration for the HR simulation. This confirms the large computational cost of building-resolving LES for urban micrometeorology. If a 20-m-resolution is used instead of a 5-m-resolution, the cost will be 256 (= $4^4$) times smaller. The elapsed time will be 2.6 min and 3.8 hours, respectively, on the supercomputer system and the desktop workstation.

The computational cost for CNN training is somewhat high. However, once the training has completed, SR prediction is not costly at all (elapsed time is negligible). Using the proposed SR simulation system, prediction with an operating cycle of less than 1 minute for urban micrometeorology with a 5-m spatial resolution is feasible on a current supercomputer system and will become feasible even on a common workstation when the computation speed becomes 100 times faster.

## 4. Conclusion

We proposed a prediction simulation system, called the SR simulation system, that consists of a physics-based prediction simulation and SR downscaling with a deep CNN. The SR downscaling maps LR images to HR ones. We focused on building-resolving simulations for urban micrometeorology. We performed building-resolving LESs using MSSG with a 5-m spatial resolution for the center of Tokyo, Japan. The training dataset for the CNN for the SR method was created from the simulation results. We showed that the proposed SR downscaling outperforms conventional interpolation methods. A spatial scaling factor of 4 is feasible for 2-m-height atmospheric temperature. This scaling factor shortens the simulation time by 256 (= $4^4$) times.

The results show that the integration of LR prediction simulations and SR downscaling can produce HR prediction results with a small computational cost and thus has the potential to realize building-resolving urban micrometeorological prediction with a real-time operating cycle. For example, prediction with an operating cycle of less than 1 mi for urban micrometeorology with a 5-m spatial resolution for a 2 km × 2 km horizontal domain is feasible on a current supercomputer system and will become feasible even on a common workstation when computation speed becomes 100 times faster.

This study focused only on temperature, which is a scalar variable. The SR downscaling performance will be different for wind, which is a vector variable. Future studies should consider other prediction variables. For more robust mapping of the physics-based simulation results, we would need a physics-informed SR method (physics SR hereafter), which accounts for the physics underlying the relation between LR and HR information. This study adopted an image SR method, not physics SR. We plan to upgrade the SR method level by level. The levels of the physics SR are shown below. We will next develop a Level 1 method, following Vandal et al. (2018).

    Level 0: Image SR method
    Level 1: Physics SR that utilizes supplemental HR information
    Level 2: Physics SR that utilizes statistical theory for mapping between LR and
        HR images
    Level 3: Physics SR that directly utilizes governing equations

**Acknowledgements**

This work was supported by the JST-Mirai Program (grant number JPMJMI18B6), Japan. The building-resolving urban simulations were performed on the Earth Simulator system of the Japan Agency for Marine-Earth Science and Technology (JAMSTEC).

**Tables**

Table 1: Datasets consisting of pairs of high-resolution (HR) and low-resolution (LR) images. Four-fifths of the image pairs were used for training and the rest were used for evaluation.

| Dataset | HR | LR | Spatial scaling factor | Number of image pairs |
|---|---|---|---|---|
| TKY.x2 | 5m | 10m | 2 | 4,300 |
| TKY.x4 | 5m | 20m | 4 | 4,300 |

Table 2: PSNR and RMSE values for the mapping between low- and high-resolution images of horizontal distribution of 2-m-height temperature.

| | | NEAREST | BICUBIC | LANCZOS | SRCNN (ImageNet) | SRCNN (MSSG) |
|---|---|---|---|---|---|---|
| PSNR [dB] | TKY.x2 | 36.8 | 38.3 | 37.7 | 40.0 | 40.8 |
| | TKY.x4 | 34.2 | 36.0 | 35.4 | 37.0 | 37.4 |
| RMSE [K] | TKY.x2 | 0.254 | 0.212 | 0.228 | 0.174 | 0.156 |
| | TKY.x4 | 0.341 | 0.276 | 0.297 | 0.245 | 0.225 |

**Figures**

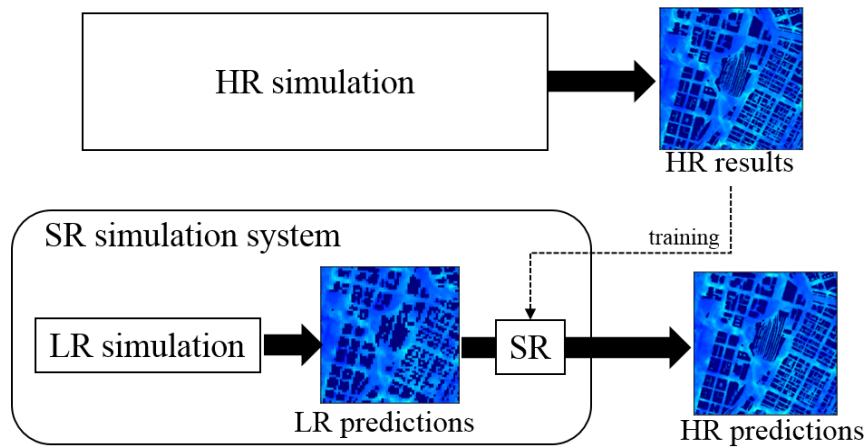

Figure 1: Concept of super-resolution (SR) simulation system for operational real-time prediction. Instead of performing high-resolution (HR) simulations to obtain HR results, low-resolution (LR) simulations are performed and the resulting LR results are converted into HR ones via SR mapping with a deep convolutional neural network (CNN) trained using the dataset obtained from HR simulations.

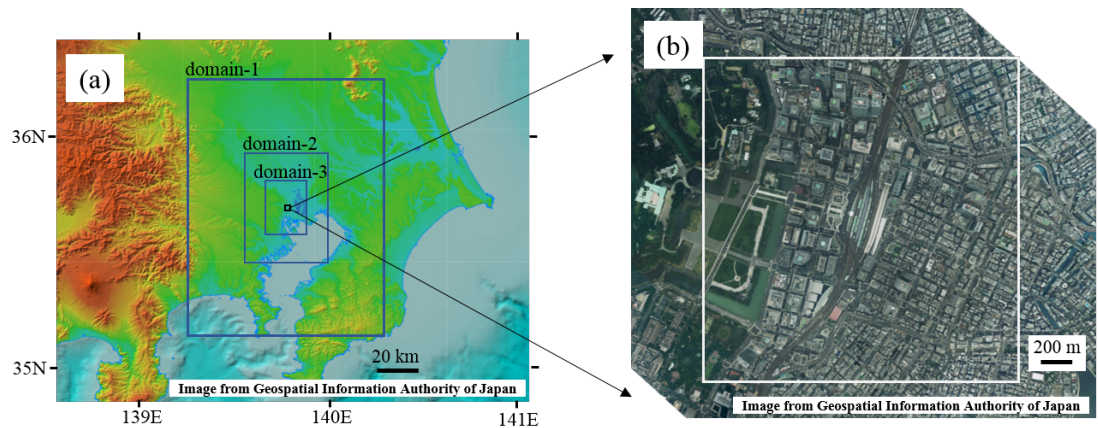

Figure 2: Computational domain. (a) mesoscale simulation domain with 3 nested domains and (b) building-resolving LES domain.

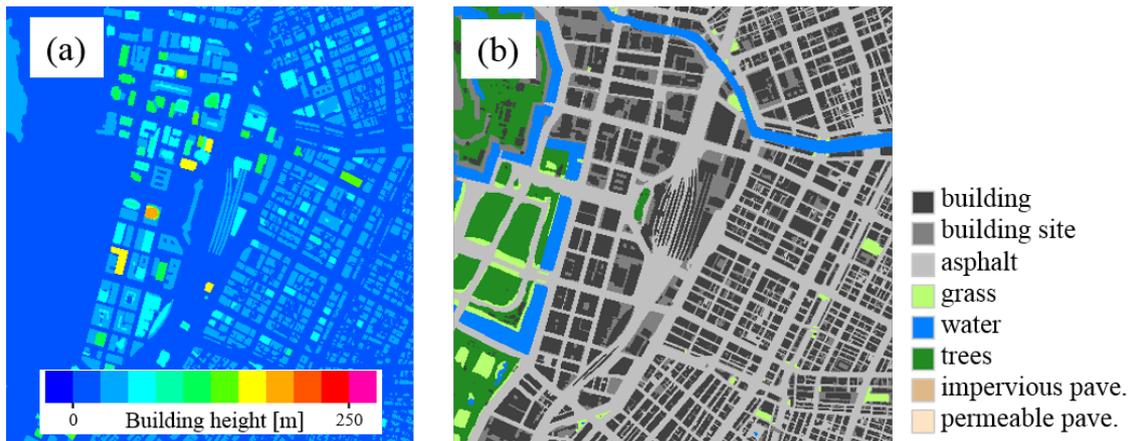

Figure 3: (a) Building height and (b) land usage in the computational domain. Tokyo Station is located in the center and the Imperial Palace is on the west (left) side of the domain.

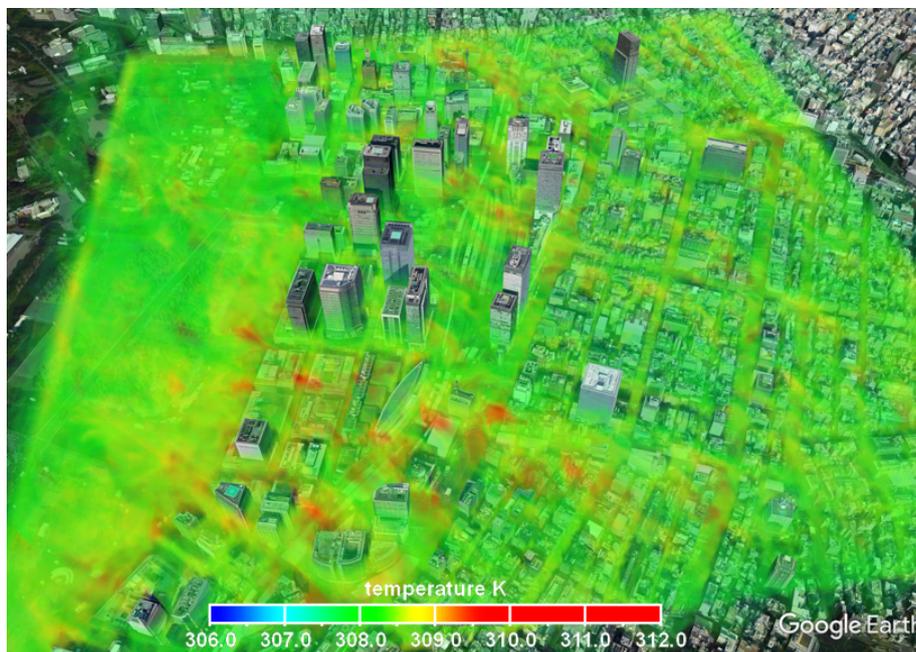

Figure 4: Three-dimensional distribution of instantaneous air temperature for 13:30 JST on 11 August, 2013.

10    *Onishi et al., SR method for real-time urban micrometeorological simulations*

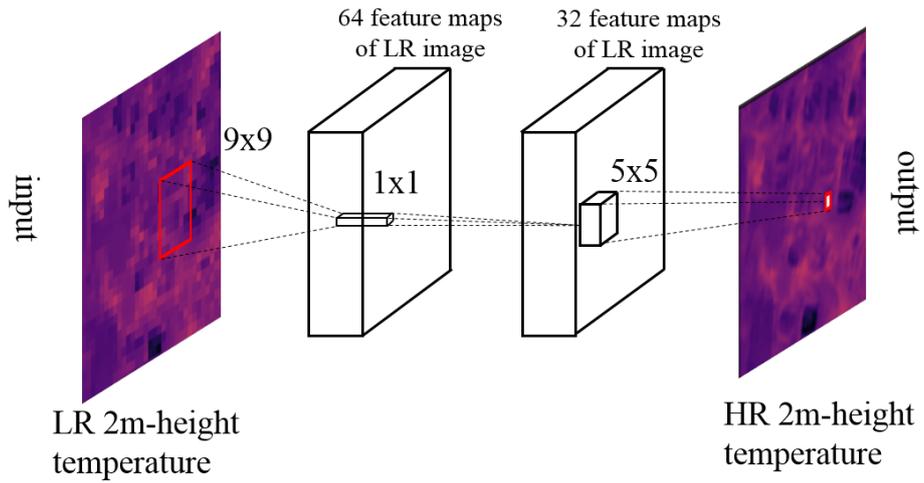

Figure 5: Illustration of the architecture of the present CNN for super-resolution simulation. The CNN learns the functional mapping between LR and HR images using patch extraction (first layer), non-linear mapping (second layer), and reconstruction (last layer).

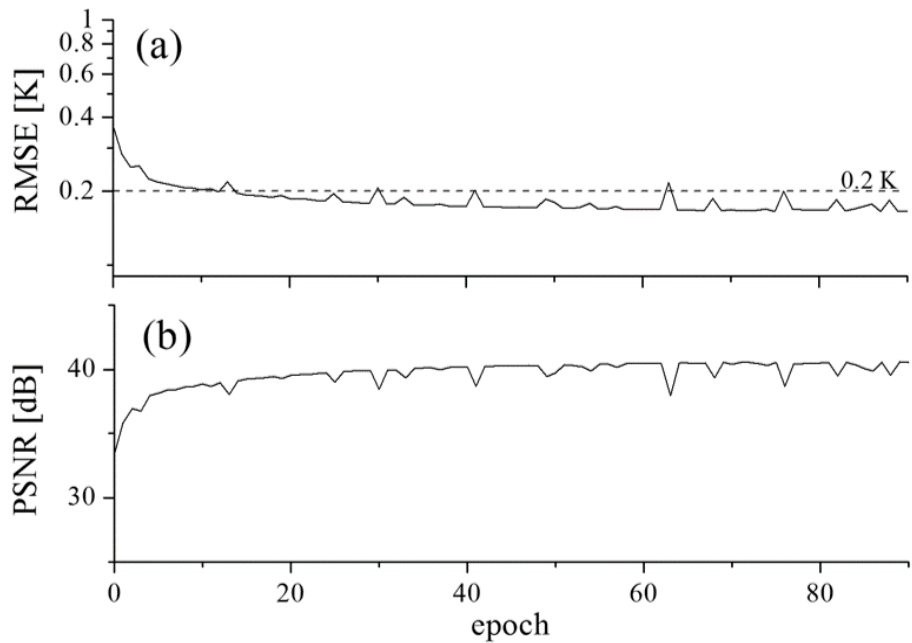

Figure 6: Learning curve of the present CNN. (a) root-mean-square-error (RMSE) and (b) peak signal-noise-ratio (PSNR). The RMSE saturates at 0.16 K, which corresponds to a PSNR of 41 dB, after 50 epochs.

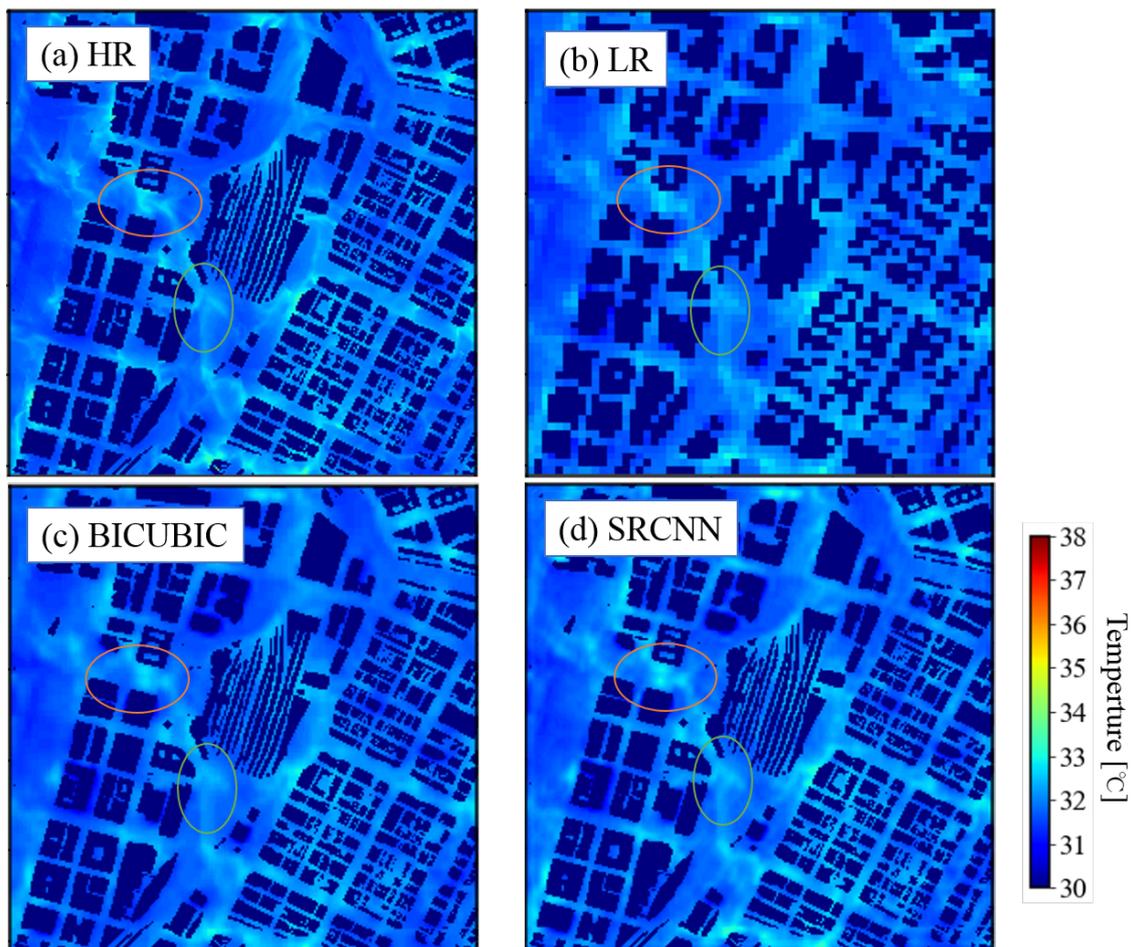

Figure 7: Example of the super-resolution downscaling for the Tokyo domain. (a) high-resolution (5-m resolution) result of 2-m height temperature (ground truth) for 15:30 JST on 25 July, 2017, (b) low-resolution (20-m resolution) result, (c) high-resolution result derived with bicubic interpolation and (d) SR-derived high-resolution mapping result.